\documentclass{amsart}
\usepackage{graphicx}
\usepackage{amsmath,amssymb}
\DeclareMathOperator*{\esssup}{\mathrm{ess\,sup}}

\newtheorem{theorem}{Theorem}%[section]
\begin{document}

\title[Stochastic evolution of a continuum particle system]{Stochastic evolution of a continuum particle system with dispersal and competition: micro- and mesoscopic description}
\author{Dmitri Finkelshtein}%
\address{Institute of Mathematics, National Academy of Sciences of Ukraine, 01601 Kiev-4, Ukraine}%
\email{dlf@imath.kiev.ua}%
\author{Yuri Kondratiev}
\address{Fakult\"at f\"ur Mathematik, Universit\"at Bielefeld, Postfach 110 131, 33501 Bielefeld, Germany}
\email{kondrat@math.uni-bielefeld.de}
\author{Yuri Kozitsky}
\address{Instytut Matematyki, Uniwersytet Marii Curie-Sklodowskiej, 20-031 Lublin, Poland}
\email{jkozi@hektor.umcs.lublin.pl}
\author{Oleksandr Kutoviy}
\address{Fakult\"at f\"ur Mathematik, Universit\"at Bielefeld, Postfach 110 131, 33501 Bielefeld, Germany}
\email{kutoviy@math.uni-bielefeld.de}
%\thanks{}%
%\subjclass{}%
%\keywords{}%

%\date{}%
%\dedicatory{}%
%\commby{}%
% ----------------------------------------------------------------
\begin{abstract}
A Markov evolution of a system of point particles in $\mathbb{R}^d$ is described
at micro-and mesoscopic levels.  The particles reproduce themselves at distant points (dispersal) and die, independently and under the influence of each other (competition). The microscopic description is based on an infinite chain of
equations for correlation functions, similar to the BBGKY hierarchy used in the Hamiltonian dynamics of continuum particle systems.  The mesoscopic description is based on a Vlasov-type kinetic equation for the particle's density obtained from the mentioned chain via a scaling procedure. The main conclusion of the microscopic theory is that the competition can prevent the system from clustering, which makes its description in terms of densities reasonable. A possible homogenization of the solutions to the kinetic equation in the long-time limit is also discussed.
\end{abstract}
\maketitle
\section{Introduction}
%\label{intro}
 Phenomenological physical theories of substances like gases, liquids, etc, are based on macroscopic observations and thus
consider these substences as continuous media.
However, already in the time of
 L. Boltzmann and M. Smoluchowski it was understood the importance of the microscopic structure of such and other similar substances. Since then a challenging problem of theoretical and mathematical physics has been to
derive the rules of the collective behavior of large systems of interacting particles from their microscopic theory, based on the so called `first principles'. Achievements in this direction laid the fundamentals of modern statistical physics; see, e.g., \cite{Dob}.
In the Hamiltonian mechanics, the motion of $N$ physical particles in $\mathbb{R}^d$ is described by a system of $2dN$ differential equations, subject to initial conditions. For $N\gg 1$, the abundance of equations, and hence of the initial conditions, makes the point-wise description meaningless since no observation could indicate at which point of the phase space the system actually is. Moreover, the point-wise description would be `too detailed' for understanding the collective behavior of the system. An alternative can be the statistical approach
which provides the possibility to link micro- and macroscopic descriptions to each other.
In this approach,  one deals with  probabilities with which such points lie in given subsets of the phase space. Then instead of `point-wise evolution' one studies the evolution of probability measures, considered now as the states of the system.
In \cite{Bog}, N. N. Bogoliubov suggested another approach, which later became popular, especially in the physical literature. Here, the evolution of probability measures is described indirectly as the evolution of the so called \textit{moment} or \textit{correlation} functions. The latter evolution is obtained from a hierarchy of equations, called now \textit{Bogoliubov} or \textit{BBGKY hierarchy} or \textit{chain}, that couples correlation functions of different order with each other,  see \cite{Dob}. The description at this level is \textit{microscopic} since one deals with coordinates of individual particles; cf. the Introduction in \cite{P}. More coarse-grained levels are \textit{meso- and macroscopic} ones. They are attained by choosing appropriate scales for space and time, see \cite{P,Spohn}. Of course, certain details of the system's behavior are lost due to such coarse-graining.
Kinetic equations provide a space-dependent mean-field-like approximate\footnote{In the physical literature, it is rather called \textit{random phase approximation.}} description of the evolution of large particle systems.
Such equations are deduced from the BBGKY chain by means of various types of scaling, cf. Section 6 in \cite{Dob} and also \cite{P,Spohn}.

Along with physics, in modern life sciences one also deals with systems consisting of
large number of interacting
 entities distributed in a continuous habitat and evolving in time.
Their collective behavior is observed at a macro-scale, and thus the
corresponding mathematical theories which explain this behavior
traditionally describe their
dynamics by means of phenomenologically (or heuristically) deduced
nonlinear equations, mostly differential or
integro-differential, involving macroscopic characteristics like
density, mobility, etc. However, this kind of
macroscopic phenomenology may often be insufficient also in life sciences, as they can collect massive experimental data of
high precision characterizing individual behavior of constituting
entities. For instance,  technical advances in animal
tracking and genotyping allow for obtaining rich individual-level
data on population dynamics, population genetics, and evolutionary
biology. However, mathematical methods are absent that would enable
one to derive the population-level consequences of such
individual-level observations.
From this and many other examples one may see how important is to elaborate a general mathematical theory in which
the rules of the collective behavior of a large number of interacting entities
can be deduced from the information about their
individual behavior and the interactions. This article is aimed at announcing some of our recent achievements in this field
based on the concepts of statistical physics mentioned above. The main principles which we follow in our study are:
\begin{itemize}
  \item The objects are large systems of entities distributed over a continuous habitat (part of $\mathbb{R}^d$, $d\geq 1$) and evolving in continuous time.
 \item The phase space of the model is the collection $\Gamma$ of possibly infinite sets of identical points -- configurations $\gamma \subset \mathbb{R}^d$.
\item In the microscopic theory, the states are probability measures on the phase space $\Gamma$, the evolution of which is assumed to be Markovian, described by Kolmogorov-type equations. The main acts of the evolution are particle's birth, death, and spatial motion -- jumps or diffusion.
    \item The evolution of states is described indirectly as the evolution of the corresponding correlation functions obeying equations similar to BBGKY chains.
    \item The mesoscopic theory is based on kinetic equations, deduced from the equations for correlation functions by means of scaling procedures, in which one scales interactions and densities, whereas time is left unscaled.
\end{itemize}

\section{The setup}
%\label{sec:1}

%\subsection{The model}

%\label{sec:2}

The object of our study in this paper is a large population of motionless entities (e.g., perennial plants) which reproduce
themselves, compete and die at random. The reproduction (dispersal) consists in random independent sending by an
entity an offspring to a distant point, which immediately
after that becomes a member of the population. Each entity can die independently with a constant rate, as well as under the influence of other population members. The latter event is interpreted as competition-caused. In theoretical biology such models were introduced and studied, e.g., in \cite{BP1,BP3,Ova}. A more detailed bibliography can be found in \cite{FKKK}.

As was suggested already in \cite{BP1}, the right mathematical
context for studying models of this kind
is the theory of random point fields in $\mathbb{R}^d$; cf. also
page 1311 in \cite{Neuhauser}. Herein, populations appear as
particle configurations constituting the set
\begin{equation} \label{C1}
 \Gamma :=\{\gamma\subset\mathbb R^d :
 |\gamma\cap K|<\infty\text{ for any bounded $K\subset\mathbb R^d$
 }\},
\end{equation}
where, for a set $A$, by $|A|$ we denote the number of points in $A$. For our model, $\Gamma$ is a phase space.
Its metric properties, as well as the main aspects of the analysis on such spaces, were studied in \cite{Tobi}.
Along with finite ones the set $\Gamma$ contains also infinite
configurations, which allows for describing `bulk' properties of a
large finite system ignoring boundary and size effects\footnote{A discussion on how infinite systems provide approximations for large
finite systems can be found in \cite{Cox,Do}.}.
If the initial configuration
$\gamma_0$ is fixed, the evolution might be described as a map
$t\mapsto \gamma_t \in \Gamma$, which in view of the random
character of the events mentioned above ought to be a random
process. However, at least so far, for the model considered here
this way can be realized only if $\gamma_0$ is finite. Hence, to describe mentioned bulk
properties one ought to go beyond the point-wise setting, as it has been done in the statistical physics
of interacting particle systems. As mentioned above, in the statistical description  states
of the system are probability measures on $\Gamma$. To characterize them
one employs {\it observables} -- appropriate functions
$F:\Gamma \rightarrow \mathbb{R}$.  As in the Heisenberg approach in quantum physics,
the evolution of states can be described via the evolution of observables. In view of the assumed Markov property, this evolution is
obtained from the equation
\begin{equation}
 \label{R2}
\frac{d}{dt} F_t = L F_t , \qquad F_t|_{t=0} = F_0, \qquad t\geq 0,
\end{equation}
in which the operator $L$ determines the model.
In our case, it is
\begin{eqnarray}\label{R20}
(LF)(\gamma) &=& \sum_{x\in \gamma}\left[ m + E^{-}
 (x, \gamma\setminus x) \right]\left[F(\gamma\setminus x) - F(\gamma) \right]\\[.2cm]
&& + \int_{\mathbb{R}^d} E^{+} (y, \gamma) \left[ F(\gamma\cup y) - F(\gamma) \right]dy, \nonumber
\end{eqnarray}
where
\begin{equation}
 \label{Ra20}
E^{\pm}(x, \gamma) := \sum_{y\in \gamma} a_{\pm} (x-y).
\end{equation}
Here $a_{\pm}\geq 0$ are suitable functions, see  (\ref{AA}) below.
As given in (\ref{R20}), $L$ is a typical `birth-and-death' generator in which the first
term corresponds to the death of the particle located at $x$
occurring (a) independently with rate $m\geq 0$, and (b) under the
influence of the other particles in $\gamma$ with rate $E^{-}(x,
\gamma\setminus x)\geq 0$. Here and in the sequel in the
corresponding context, we treat each $x\in \mathbb{R}^d$ also as a
single-point configuration $\{x\}$. That is, if $x$ belongs to $\gamma$ (resp. $y$ does not), by
$\gamma\setminus x$ (resp. $\gamma \cup y$) we mean the configuration which is obtained from $\gamma$ by removing
$x$ (resp. by adding $y$).
The
second term in (\ref{R20}) describes the birth of a particle at
$y\in \mathbb{R}^d$ given by the whole configuration $\gamma$ with
rate $E^{+}(y, \gamma)\geq 0$. The model described
by (\ref{R20}) is called a {\it spatial logistic model
(SLM).} A~particular case of SLM is the continuous contact model
\cite{KKP,KS} where $a_{-} \equiv 0$, and hence the competition is
absent.

The sums in (\ref{R20}) and (\ref{Ra20}) may contain infinite number of summands and hence may diverge.
This means that the direct study of the corresponding equation (\ref{R2}) for our model is rather problematic.
Following the way suggested by N. N. Bogoliubov, the evolution of states  will be studied via the evolution of the corresponding
correlation functions $k_0 \mapsto k_t$. By a correlation function $k$ we mean an infinite collection of
symmetric positive functions $k^{(n)}:(\mathbb{R}^d)^n \to \mathbb{R}$,
$n\in \mathbb{N}_0$, which determine the state in a certain way, see \cite{FKKK,Dima,DimaN,DimaNN} for more detail.
Such $k$ can also be viewed as infinite-dimensional (Fock-type) vectors.
Note that $k^{(0)} = 1$ and $k^{(1)}$ is the particle density. For the Poisson measure with intensity $\varkappa >0$,
 one has $k^{(n)} (x_1 , \dots , x_n)= \varkappa^n$, $n\in \mathbb{N}_0$. We say that a correlation function $k$, and hence the corresponding state, are \textit{sub-Poissonian} if
\begin{equation}
  \label{sp}
 k^{(n)} (x_1 , \dots , x_n) \leq C^n,
\end{equation}
which holds for some $C>0$, all $n\in \mathbb{N}_0$, and (Lebesgue) almost all $(x_1, \dots , x_n)$.
Then a sub-Poissonian state is, in a way, similar to the Poissonian state in which the particles are independently
scattered over $\mathbb{R}^d$. At the same time, the increase of $k^{(n)}$ with $n$ as $n!$ (see \eqref{nfactorial}
below), may correspond to the formation of \textit{clusters}.
The
evolution $k_0\mapsto k_t$ is obtained from the equation
\begin{equation}
 \label{R4}
\frac{d}{dt} k_t = L^\Delta k_t, \qquad k_t|_{t=0} = k_0,
\end{equation}
which, in fact, is a chain of equations for particular $k_t^{(n)}$, analogous to the Bogoliubov hierarchy for physical particles.
The generator $L^\Delta$ is constructed from $L$ as in (\ref{R20}) by a certain procedure, see \cite{Dima,DimaNN}. In particular, the first equations in the chain
(\ref{R4}) have the following form: $d k^{(0)}_t / dt = 0$ and
\begin{equation}
  \label{eeq1}
 \frac{d}{dt} k_t^{(1)} (x) = - m k^{(1)}_t(x) -  \int_{\mathbb{R}^d} a_{-} (x-y) k^{(2)}_t(x,y)dy + \int_{\mathbb{R}^d} a_{+} (x-y) k^{(1)}_t(y)dy .
\end{equation}
The RHS of the equation with $d k^{(2)}_t / dt$ contains $k^{(n)}_t$ with $n=1,2,3$, etc.
Theoretical biologists try to solve chains like (\ref{R4}) by decoupling them; cf. \cite{Mu}.
In the simplest version,
one sets
\begin{equation*}
 % \label{e11}
k^{(2)}(x, y) \simeq k^{(1)} (x )  k^{(1)} (y),
\end{equation*}
which is equivalent to neglecting spatial pair correlations. This turns (\ref{eeq1}) into the following nonlinear equation
\begin{equation}
  \label{eeq2}
 \frac{d}{dt} k_t^{(1)} (x) = - m k^{(1)}_t(x) -  k^{(1)}_t (x)\int_{\mathbb{R}^d} a_{-} (x-y) k^{(1)}_t(y)dy + \int_{\mathbb{R}^d} a_{+} (x-y) k^{(1)}_t(y)dy ,
\end{equation}
which, in fact, is a kinetic equation for our model.
Its eventual solution provides a mean-field-like approximation of the evolution of the density. Note that the question of
whether the evolution of states preserves sub-Poissonicity, and hence clusters do not occur, can be answered only
by studying the whole chain in (\ref{R4}). For the contact model, which one obtains from
(\ref{R20}) by setting $a_{-} \equiv 0$, it is known \cite{Dima} that
\begin{equation}\label{nfactorial}
\mathrm{const}\cdot n!\, c_t^n\leq k_t^{(n)} (x_1 , \dots , x_n)
\leq \mathrm{const}\cdot n!\, C_t^n,
\end{equation}
where the left-hand inequality holds if all $x_i$ belong to a ball
of small enough radius. Hence, in spite of the fact that $C_t \to 0$
as $t\to +\infty$ if the mortality dominates the dispersal, $k_t$ are definitely not sub-Poissonian if
$a_{-} \equiv 0$.

In this article we address the following questions:
\begin{enumerate}
  \item Can the competition in (\ref{R20}) provide that the problem (\ref{R4}) has a solution that obeys (\ref{sp})?
      \item Can one deduce (\ref{eeq2}) rigorously; e.g., by a kind of scaling procedure, which would allow for error control?
      \item What can be said of solutions to (\ref{eeq2})  in that case?
\end{enumerate}
The answers are given in the form of theorems stated below. The proof of Theorem \ref{1tm} can be found in \cite{FKKK}; the remaining
theorems will be proven in our work under preparation.

\section{The results}
%\label{3S}
We assume that the competition and dispersal rates are additive, cf. (\ref{Ra20}), and the corresponding \textit{kernels}
$a_{\pm}$ are symmetric, essentially bounded, and integrable. That is,
\begin{equation}
 \label{AA}
a_{\pm} \in L^1(\mathbb{R}^d)\cap L^\infty(\mathbb{R}^d), \qquad a_{\pm}(x) = a_{\pm}(-x) \geq 0,
\end{equation}
and thus we set
\begin{equation}
 \label{14A}
\langle{a}_{\pm}\rangle = \int_{\mathbb{R}^d} a_{\pm} (x)dx,\qquad \|a_{\pm} \|= \esssup_{x\in \mathbb{R}^d} a_{\pm}(x).
\end{equation}
Let $\Gamma_0$ be the subset of $\Gamma$ as in (\ref{C1})
consisting of finite configurations only. For $\gamma \in \Gamma_0$, by $|\gamma|$ we denote the number of points in $\gamma$.
One observes that, for $\gamma \in \Gamma_0$, the sums in (\ref{Ra20}) and in the expressions below are well-defined. Set
\begin{equation*}
% \label{15A}
E^{\pm} (\gamma) = \sum_{x\in \gamma}E^{\pm} (x,\gamma\setminus x) = \sum_{x\in \gamma} \sum_{y\in \gamma\setminus x}a_{\pm} (x-y), \quad
\gamma \in \Gamma_0 .
\end{equation*}
%By a function $g:\Gamma_0 \to \mathbb{R}$ we then mean a collection of symmetric functions $g^{(n)}: (\mathbb{R}^d)^n \to \mathbb{R}$, $n\in %\mathbb{N}_0$, where $g^{(0)}$ is just a constant function.
%For such $g$, we write
%\begin{equation}
 % \label{14}
%\int_{\Gamma_0} g(\gamma) \lambda (d \gamma) := \sum_{n=0}^\infty   \frac{1}{n!} \int_{(\mathbb{R}^d)^n} g^{(n)} (x_1 , \dots , x_n) dx_1 \cdots %dx_n,
%\end{equation}
%and assume that the integral in the left-hand side exists if all the integrals in the right-hand side exist and that the series converges.
Then the operator in (\ref{R4}) acts as follows
\begin{eqnarray}
  \label{15}
(L^\Delta k)(\gamma) & = & - (m |\gamma| + E^{-}(\gamma)) k(\gamma) + \sum_{x\in \gamma} E^{+}(x, \gamma\setminus x) k(\gamma\setminus x) \\[.2cm]
& - & \int_{\mathbb{R}^d} E^{-}(x, \gamma) k (\gamma \cup x) dx + \int_{\mathbb{R}^d} \sum_{x\in \gamma}a_{+} (x-y) k(\gamma\setminus x \cup y) dy, \nonumber
\end{eqnarray}
where $\gamma \in \Gamma_0$. For a one-point $\gamma =\{ x\}$, one readily derives from (\ref{15}) the right-hand side of (\ref{eeq1}).

So far, (\ref{15}) is just a formal expression. Then our next aim is to introduce a Banach space, in which $L^\Delta$ can be defined as a linear operator, and hence   (\ref{R4}) can be studied. For $k:\Gamma_0 \to \mathbb{R}$, let us assume that each $k^{(n)}: (\mathbb{R}^d)\to \mathbb{R}$ is essentially bounded, and then, for $\alpha \in \mathbb{R}$,  set
\begin{equation}
  \label{16}
\|k\|_\alpha = \sup_{n\in \mathbb{N}_0} e^{\alpha n} q_n (k),
\end{equation}
where $q_0 (k) = |k^{(0)}|$, and
\[
q_n (k) := \esssup_{(x_1 , \dots , x_n )\in  (\mathbb{R}^d)^n} |k^{(n)}(x_1 , \dots , x_n )|, \qquad n\in \mathbb{N}.
\]
Afterwards, we set
\begin{equation}
  \label{17}
 \mathcal{K}_\alpha = \{ k: \Gamma_0 \to \mathbb{R} : \|k\|_\alpha< \infty\}.
\end{equation}
That is, $\mathcal{K}_\alpha$ is an infinite dimensional real vector space equipped with norm (\ref{16}), complete with respect to this norm.
For $\alpha'' < \alpha'$, we have $\|k\|_{\alpha''} \leq \|k\|_{\alpha'}$; and hence,
\begin{equation}
  \label{18}
 \mathcal{K}_{\alpha'} \subset \mathcal{K}_{\alpha''}, \qquad {\rm for} \ \ \alpha'' < \alpha'.
\end{equation}
That is, the bigger is $\alpha$, the `smaller' is the corresponding space (\ref{17}).
For $k\in \mathcal{K}_\alpha$, by (\ref{16}) we readily have $q_n(k) \leq \|k\|_\alpha \exp( - \alpha n)$, which yields that $k$ is sub-Poissonian; cf. (\ref{sp}). Thus, for each $\alpha\in \mathbb{R}$, the space (\ref{17}) contains only sub-Poissonian $k$, and each such a $k$ is contained in some $\mathcal{K}_\alpha$. Therefore, the collection of spaces (\ref{17}) with all possible
$\alpha \in \mathbb{R}$, cf. (\ref{18}), provides an appropriate framework for describing the evolution of sub-Poissonian
correlation functions.

Now let us turn to defining the operator (\ref{15}). First of all, we note that, for $k\in \mathcal{K}_\alpha$, $L^\Delta k$ need not be in the same $\mathcal{K}_\alpha$, which means that $L^\Delta$, as a map from $\mathcal{K}_\alpha$ into itself, cannot be defined on the whole $\mathcal{K}_\alpha$. On the other hand, it is possible to obtain the following estimate, see Section 4 in \cite{FKKK},
\begin{eqnarray*}
\|L^\Delta k \|_\alpha \leq \psi (\alpha' - \alpha) \|k\|_{\alpha'}, \qquad \alpha'> \alpha,
\end{eqnarray*}
where $\psi: (0,+\infty) \to (0,+\infty)$ is a certain function such that $\psi(\tau) \to +\infty$
as $\tau \to 0^+$. That is $L^\Delta$ acts as a bounded operator to $\mathcal{K}_\alpha$ from any smaller $\mathcal{K}_{\alpha'}$.
Then the domain of $L^\Delta$ in $\mathcal{K}_\alpha$ is defined to be the set of all those elements of this space which belong to smaller spaces $\mathcal{K}_{\alpha'}$. That is
\begin{equation*}
 % \label{19}
\mathcal{D}_\alpha = \bigcup_{\alpha' : \alpha' > \alpha} \mathcal{K}_{\alpha'}.
\end{equation*}
Now we can make precise what do we mean by solving (\ref{R4}). Namely, for $0< T \leq +\infty$, a map $[0,T) \ni t \mapsto k_t \in \mathcal{K}_\alpha$ is said to be a \textit{classical solution}\footnote{In infinite dimensional Banach spaces, there can be solutions in  weaker senses.} to the problem (\ref{R4}) in space $\mathcal{K}_\alpha$ on time interval $[0,T)$ if: (a) it is continuously differentiable on $(0,T)$; (b) $k_t\in \mathcal{D}_\alpha$ for all $t\in (0,T)$; (c) it obeys both the equation and the initial condition in (\ref{R4}).

Let us take some ${\alpha}_* \in \mathbb{R}$, and then any ${\alpha}^* > {\alpha}_*$, and then set
\begin{equation}
  \label{20}
  T_* = \frac{\alpha^* - \alpha_*}{\langle a_{+} \rangle + \langle a_{-} \rangle e^{-\alpha_*}},
\end{equation}
where $\langle a_{\pm} \rangle$ are the same as in (\ref{14A}). Our first result is given in the next statement.
\begin{theorem}
  \label{1tm}
Suppose that  there exists $\theta >0$ such that, for almost all $x\in \mathbb{R}^d$,
\begin{equation}
  \label{21}
a_{+} (x) \leq \theta a_{-} (x).
\end{equation}
Then, for each $\alpha^*\in \mathbb{R}$ such that
\begin{equation}
  \label{22}
\theta e^{\alpha^*} < 1,
\end{equation}
and any $\alpha_* < \alpha^*$, the problem (\ref{R4}) with $k_0 \in \mathcal{K}_{\alpha^*}$ has a unique classical solution in $\mathcal{K}_{\alpha_*}$ on time interval $[0,T_*)$, with $T_*$ given in (\ref{20}).
\end{theorem}
Let us make some comments on this result. First of all we note that the solution
to (\ref{R4}) is sub-Poissonian, as hence clusters do not occur,  if the initial $k_0$ shares this property. Second, the solution lies in a `bigger' space
than $k_0$ does. The choice of this initial space is restricted by (\ref{22}). For a fixed $\alpha^*$, $T_*$  has a unique maximum as a function of $\alpha_*$, that means that there exists an optimal choice of $\alpha_*$, for which
the time interval $[0,T_*)$ has maximal length.
In (\ref{21}) we, in fact, assume that the dispersal is dominated by the competition.
The essence of this condition can be formulated as follows: each particle can `kill' its offspring and can be `killed' by it.

Now let us turn to the mesoscopic theory, which we base on a Vlasov-type scaling; cf. Section 6 in \cite{Do}. According to the general scheme developed in \cite{DimaN}, this amounts to passing to the scale at which the microscopic density gets of order $\varepsilon^{-1}$ and the  interaction of order $\varepsilon$, where
$\varepsilon \in (0,1]$ is a scaling parameter. This means that the initial correlation function $k_0(\gamma)$ gets of order
$\varepsilon^{-|\gamma|}$; and hence, one assumes that
\begin{equation}
  \label{23}
 \varepsilon^{|\gamma|} k_0(\gamma) \to r_0 (\gamma), \qquad {\rm as} \ \ \varepsilon \to 0^+, \qquad \gamma \in \Gamma_0,
\end{equation}
where $r_0$ is a correlation function. Additionally, one assumes that also $k_t$ for $t>0$, have the same divergence, and hence should be
`renormalized' as in (\ref{23}). The interaction in $L^\Delta$ is represented by the competition kernel $a_{-}$ which thus  should be replaced by $\varepsilon a_{-}$ (weak interaction limit). Let $L^\Delta_\varepsilon$ be the operator which we obtain by such a replacement. Then we set, cf. (\ref{23}),
\begin{equation}
  \label{24}
k^{(\varepsilon)}_{t, {\rm ren}} (\gamma) = \varepsilon^{|\gamma|} k_t(\gamma), \qquad (L^\Delta_{\varepsilon, {\rm ren}} k)(\gamma) =
\varepsilon^{|\gamma|} (L^\Delta_\varepsilon k) (\gamma).
\end{equation}
Thereafter, the problem (\ref{R4}) turns into the following
\begin{equation}
  \label{25}
 \frac{d}{dt}  k^{(\varepsilon)}_{t, {\rm ren}} = L^\Delta_{\varepsilon, {\rm ren}} k^{(\varepsilon)}_{t, {\rm ren}}, \qquad  k^{(\varepsilon)}_{t, {\rm ren}}|_{t=0} =  k^{(\varepsilon)}_{0, {\rm ren}}.
\end{equation}
The expression for $L^\Delta_{\varepsilon, {\rm ren}}$ can be calculated from (\ref{24}) explicitly, which yields it in the form
\begin{equation}
  \label{26}
 L^\Delta_{\varepsilon, {\rm ren}} = V + \varepsilon B,
\end{equation}
where, cf. (\ref{15}),
\begin{eqnarray}
  \label{27}
(V k)(\gamma) & = & - m k(\gamma)  - \int_{\mathbb{R}^d} E^{-}(x, \gamma) k (\gamma \cup x) dx\\[.1cm]& + & \int_{\mathbb{R}^d} \sum_{x\in \gamma}a_{+} (x-y) k(\gamma\setminus x \cup y) dy, \nonumber
\end{eqnarray}
and
\[
(B k)(\gamma) = - E^{-} (\gamma) k(\gamma) + \sum_{x\in \gamma} E^{+}(x, \gamma\setminus x) k(\gamma\setminus x).
\]
Note that $L^\Delta_{\varepsilon, {\rm ren}}|_{\varepsilon = 1} = L^\Delta$.
For each $\varepsilon \in (0,1]$,
(\ref{25}) has a unique classical solution in the same space and on the same time
interval as (\ref{R4}) does. This can be proven exactly as Theorem \ref{1tm}. For $r_0$ as in (\ref{23}) and $V$ as in (\ref{27}), let us consider
\begin{equation}
  \label{28}
   \frac{d}{dt} r_t = V r_t, \qquad r_t|_{t=0} = r_0.
\end{equation}
For this problem, one can probe an analog of Theorem \ref{1tm}, even without assuming (\ref{21})
because the latter  condition relates to the operator $B$, cf. (\ref{26}).
\begin{theorem}
  \label{2tm}
Let $\alpha^*$, $\alpha_*$, and $T_*$ be as in Theorem \ref{1tm}, and let $k^{(\varepsilon)}_{t, {\rm ren}}$ (resp. $r_t$) be solutions to
(\ref{24}) (resp. (\ref{28})) in $\mathcal{K}_{\alpha_*}$ on time interval $[0,T_*)$, with the initial conditions $k^{(\varepsilon)}_{t, {\rm ren}}|_{t=0} = r_t|_{t=0} = r_0 \in \mathcal{K}_{\alpha^*}$. Then, for every positive $T<T_*$, it follows that
\[
\sup_{t \in [0,T]} \|k^{(\varepsilon)}_{t, {\rm ren}} - r_t\|_{\alpha_*}  \to 0, \qquad {\rm as} \ \
\varepsilon \to 0^+.
\]
\end{theorem}
The equation in (\ref{28}) is, in fact, a chain of equations for $r^{(n)}$, $n\in\mathbb{N}_0$. It is called a \textit{Vlasov hierarchy}.
Theorem \ref{2tm} states that, in the scaling limit $\varepsilon \to 0^+$, the solutions to (\ref{25}) converge in $\mathcal{K}_{\alpha_*}$
to the solution to the Vlasov hierarchy, uniformy on closed subintervals of $[0,T_*)$. The reason to pass from $k_t = k^{(\varepsilon)}_{t, {\rm ren}}|_{\varepsilon =1}$ to $r_t = k^{(\varepsilon)}_{t, {\rm ren}}|_{\varepsilon=0}$ can be seen from the following arguments.
Suppose that $r_0$ in (\ref{28}) is such that $r^{(0)}_0 =1$ and
\begin{equation*}
 % \label{29}
  r_0^{(n)} (x_1 , \dots , x_n) = \varrho_0 (x_1) \cdots \varrho_0(x_n), \qquad n\in \mathbb{N},
\end{equation*}
for some positive $\varrho_0\in L^\infty (\mathbb{R}^d)$. In other words, we assume that $r_0$ is the correlation function of a heterogeneous Poisson measure, with density function $\varrho_0$.  Let  us now seek the solution to (\ref{28}) also as the product of certain unknown $\varrho_t$. The peculiarity of $V$ as in (\ref{27}) is that it is possible to do, and that $\varrho_t$ can be found from the following equation
\begin{equation}
  \label{30}
 \frac{d}{dt} \varrho_t (x) = - m \varrho_t(x) -  \varrho_t (x)\int_{\mathbb{R}^d} a_{-} (x-y) \varrho_t(y)dy + \int_{\mathbb{R}^d} a_{+} (x-y) \varrho_t(y)dy ,
\end{equation}
with the initial condition $\varrho_t|_{t=0} = \varrho_0$, which is exactly the kinetic equation (\ref{eeq2}). In particular, the passage from $\varepsilon =1$ to
$\varepsilon =0$
has led us from (\ref{eeq1}) to (\ref{eeq2}). Since we know that the Vlasov hierarchy (\ref{28}) has a unique solution, then
this solution has the form $r_t^{(n)} (x_1 , \dots , x_n) = \varrho_t(x_1)\cdots \varrho_t(x_n)$, $n\in \mathbb{N}$, whenever
(\ref{30}) has a unique solution in $L^\infty(\mathbb{R}^d)$ on time interval $[0,T_*)$, which obeys the condition
\begin{equation*}
 % \label{31}
  \| \varrho_t \|_{L^\infty(\mathbb{R}^d)} \leq e^{-\alpha_*}.
\end{equation*}
The latter is to guarantee that
$r_t$ lies in $\mathcal{K}_{\alpha_*}$; cf. (\ref{16}) and (\ref{17}).

Our next problem is to study the solvability
of (\ref{30}). For some technical reasons, it is more convenient to solve it in the space $C_{\rm b} (\mathbb{R}^d)$ of all bounded continuous functions $\phi: \mathbb{R}^d \to \mathbb{R}$. We equip this space with the norm
\[
\|\phi \|= \sup_{x \in \mathbb{R}^d} |\phi (x) |,
\]
which turns it into a Banach space, that can be isomorphically embedded into  $L^\infty(\mathbb{R}^d)$.
We say that $\phi \in  C_{\rm b} (\mathbb{R}^d)$ is \textit{positive} if $\phi(x) \geq 0$ for all $x\in \mathbb{R}^d$.
Our first result in this domain
is the following statement.
\begin{theorem}
  \label{3tm}
For each positive $\varrho_0 \in   C_{\rm b} (\mathbb{R}^d)$, the problem (\ref{30}) with
the initial condition $\varrho_t|_{t=0} = \varrho_0$ has a unique classical positive solution $\varrho_t \in  C_{\rm b} (\mathbb{R}^d)$ on time interval $[0, +\infty)$.
\end{theorem}
Now let $\varrho_0 \in   C_{\rm b} (\mathbb{R}^d)$ be constant as a function of $x$. Then the solution as in Theorem \ref{3tm} will also be independent of $x$ and can be obtained explicitly. Set $\varrho_t (x) \equiv u_t$. Then $u_t$ has to solve the following
\begin{equation}
  \label{32}
 \frac{d}{dt} u_t = (\langle a_{+} \rangle - m) u_t - \langle a_{-} \rangle u_t^2,
\end{equation}
see (\ref{14A}), which is a Bernoulli equation. For $m > \langle a_{+} \rangle$, its solution decays to zero exponentially as $t \to +\infty$. For
$m = \langle a_{+} \rangle$, the solution is $u_t = u_0/(1 +  \langle a_{-} \rangle u_0 t)$, and hence also decays to zero as $t \to +\infty$.
For $m < \langle a_{+} \rangle$, we set
\begin{equation*}
%  \label{33}
  q = \frac{ \langle a_{+} \rangle - m}{\langle a_{-} \rangle}.
\end{equation*}
In this case the solution to (\ref{32}) has the form
\begin{equation}
  \label{34}
  u_t = \frac{u_0 q}{u_0 + (q-u_0) \exp(- q \langle a_{-} \rangle t)},
\end{equation}
which, in particular, means that $u_t \to q$ as $t \to +\infty$.
\begin{theorem}
 % \label{4tm}
Suppose that $q>0$ and, for almost all $x\in \mathbb{R}^d$, the following holds
\begin{equation}
  \label{35}
 \frac{a_{+} (x)}{\langle a_{+} \rangle} \geq \left(1 - \frac{m}{\langle a_{+} \rangle} \right) \frac{a_{-} (x)}{\langle a_{-} \rangle}.
\end{equation}
Let also the initial condition $\varrho_0 \in C_{\rm b}(\mathbb{R}^d)$ to the problem (\ref{30}) obeys
\begin{equation*}
 % \label{36}
  0< \delta \leq \varrho_0 (x) < q,
\end{equation*}
which holds for some $\delta$ and all $x\in \mathbb{R}^d$. Then, for each $x\in \mathbb{R}^d$ and $t>0$, the solution as in Theorem \ref{3tm} obeys the bounds $u_t \leq \varrho_t(x) < q$, where $u_t$ is given in (\ref{34}) with $u_0=\delta$. Hence $\varrho_t (x) \to q$ point-wise as $t\to +\infty$.
\end{theorem}
The result just obtained can be interpreted as an asymptotic homogenization of the density. Let us add some comments to (\ref{35}).
If $a_{+} (x)= \theta a_{-}(x)$, for some $\theta>0$ and almost all $x$, cf. (\ref{21}), then (\ref{35}) holds for all $m\in [0, \langle a_{+} \rangle)$. Note that the case of $a_{+} (x)=  a_{-}(x)$ was studied in \cite{Perthame}. In population biology, competition often has the range shorter than that of dispersal. In such a case, the mentioned homogenization
occurs at nonzero mortality $m$. For instance, let $a_{+} = \alpha \mathbb{I}_R$ and $a_{-} = \beta \mathbb{I}_r$ for some positive $\alpha$, $\beta$, $r$, and $R$, and also $R\geq r$. Here $\mathbb{I}_r(x) = 1$ if $|x| \leq r$ and $\mathbb{I}_r(x) = 0$ otherwise. Then (\ref{35}) holds if
\[
1 - \frac{m}{\langle a_{+} \rangle} \leq \left(\frac{r}{R} \right)^d.
\]

\section*{Acknowledgment}
This paper was supported by the DFG through the SFB 701
``Spektrale Strukturen und Topologische Methoden in der Mathematik''
and through the research project 436 POL 125/0-1.

%\label{s3}
%\begin{figure}
%\resizebox{0.75\columnwidth}{!}{%
%  \includegraphics{fig1.eps} }
%\caption{Please write your figure caption here.}
%\end{figure}

\end{document}